\documentclass[%
 reprint,
 nofootinbib,
 amsmath,amssymb,
 aps,
]{revtex4-1}

\usepackage{lipsum}
\usepackage{balance}
\usepackage{hyperref}
\usepackage{appendix}
\usepackage[user,titleref]{zref}
\usepackage[nameinlink,capitalize]{cleveref}
\usepackage{braket}
\usepackage{natbib}
\usepackage{ulem}
\usepackage{subfigure}
\usepackage{tikz}
\usepackage{graphicx}
\usepackage{dcolumn}
\usepackage{bm}

\begin{document}

\preprint{APS/123-QED}

\title{Searching for ultralight bosons with supermassive black hole ringdown}

\author{Adrian Ka-Wai Chung}
\email{ka-wai.chung@ligo.org}
\affiliation{%
 Theoretical Particle Physics and Cosmology Group, Department of Physics, King’s College London, University of London, Strand, London, WC2R 2LS, United Kingdom
}%

\author{Joseph Gais}
\email{1155138494@link.cuhk.edu.hk}
\affiliation{%
Department of Physics, The Chinese University of Hong Kong, Shatin, N.T., Hong Kong.
}%

\author{Mark Ho-Yeuk Cheung}
\email{hcheung5@jhu.edu}
\affiliation{%
Department of Physics and Astronomy, Johns Hopkins University, 3400 N. Charles Street, Baltimore, Maryland 21218, USA
}%
\affiliation{%
Department of Physics, The Chinese University of Hong Kong, Shatin, N.T., Hong Kong.
}%

\author{Tjonnie G.F. Li}
\email{tgfli@cuhk.edu.hk}
\affiliation{%
 Department of Physics, The Chinese University of Hong Kong, Shatin, N.T., Hong Kong
}%
\affiliation{%
Institute for Theoretical Physics, KU Leuven, Celestijnenlaan 200D, B-3001 Leuven, Belgium
}%
\affiliation{%
Department of Electrical Engineering (ESAT), KU Leuven, Kasteelpark Arenberg 10, B-3001 Leuven, Belgium
}%

\newcommand{\mc}[1]{{\color{red}[Mark: #1]}}
\newcommand{\mcst}[1]{{\color{red}[Mark: (remove) \sout{#1}]}}
\newcommand{\acadd}[1]{{\color{blue}[Adrian: (add) \st{#1}]}}
\newcommand{\acrm}[1]{{\color{blue}[Adrian: (remove) \sout{#1}]}}
\newcommand{\jcst}[1]{{\color{green}[Joey: (remove) \sout{#1}]}}
\newcommand{\jc}[1]{{\color{green}[Joey: #1]}}

\begin{abstract}
One class of competitive candidates for dark matter is ultralight bosons.
If they exist, these bosons may form long-lived bosonic clouds surrounding rotating black holes via superradiant instabilities, acting as sources of gravity and affecting the propagation of gravitational waves around the host black hole.
During extreme-mass-ratio inspirals, the bosonic clouds will survive the inspiral phase and can affect the quasinormal-mode frequencies of the perturbed black-hole-bosonic-cloud system.
In this work, we compute the shifts of gravitational quasinormal-mode frequencies of a rotating black hole due to the presence of a surrounding bosonic cloud.
We then perform a mock analysis on simulated Laser Interferometer Space Antenna observational data containing injected ringdown signals from supermassive black holes with and without a bosonic cloud.
We find that with less than an hour of observational data of the ringdown phase of nearby supermassive black holes such as Sagittarius A* and M32, we can rule out or confirm the existence of cloud-forming ultralight bosons of mass $ \sim 10^{-17} \rm eV$.
\end{abstract}

\pacs{Valid PACS appear here}
\maketitle

\section{Introduction} 
Ultralight bosons are particles with mass $\ll 1 \, \rm eV$ \cite{peccei1977constraints, weinberg1978new, wilczek1978problem}.
This class of particles has been a promising dark matter candidate which may address a number of outstanding problems in fundamental physics ranging from particle physics \cite{bertone2005particle, arvanitaki2010string, arvanitaki2011exploring, arvanitaki2015searching} to cosmology \cite{cosmology_01,cosmology_02, cosmology_03, cosmology_04, cosmology_05}.
If they exist, they will form condensates around a rotating black hole of comparable size to their Compton wavelength \cite{zel1971generation, press1972floating, brito2020superradiance, palomba2019direct}. 
The condensate will grow exponentially by extracting rotational energy from the black hole, settling into a ``bosonic cloud".
Although the cloud will then decay by emitting gravitational radiation, the timescale of such decay is very long, so bosonic clouds could exist on cosmological timescales \cite{okawa2014black}.

Some methods have been proposed to search for ultralight bosons around black holes, such as by measuring the dephasing of binary mergers due to dynamical friction \cite{eda2013new, annulli2020response, kavanagh2020detecting, berti2019ultralight}, quasimonochromatic radiation from the bosonic cloud \cite{palomba2019direct, okawa2014black, tsukada2019first, tsukada2020modeling}, holes in the spin-mass plane of black hole populations due to the superradiant energy extraction of bosonic clouds \cite{ng2019searching, ng2020constraints}, and other signatures during the inspiral phase of binary mergers \cite{hannuksela2019probing, baumann2019probing, macedo2013into}.

Most of these methods are concerned with the effects of ultralight bosons on the inspiral phase of binary black hole mergers. 
However, because of the bosonic cloud, the spacetime around the black hole is no longer vacuum. 
Matter effects due to the presence of this cloud have been demonstrated for the inspiral phase of binary mergers \cite{choudhary2020gravitational, 2020PhRvD.101h4055J, barausse2014can}, but we demonstrate that the cloud also affects the ringdown phase, during which a perturbed black hole relaxes into a stationary black hole by emitting gravitational waves (GWs) in a discrete set of complex quasinormal modes frequencies (QNMFs).

Gravitational perturbations around a rotating black hole are governed by the Teukolsky equation.
Because of the gravity of the bosonic cloud around a black hole, an additional effective potential peaked at $r \sim \mu^{-1} > M $ arises in the Teukolsky equation if $M \mu < 1 $.
As the ringdown GWs propagate away from the black hole, it will first reach the usual effective potential peaked at the angular momentum barrier of the black hole.
If a bosonic cloud exists, as the waves propagate further outward, they will also encounter the cloud's potential. 
This additional effective potential can modify the quasinormal-modes of the system, resulting in an altered ringdown waveform upon detection.

In this work, we envisage the ringdown GW signals emitted by the closest supermassive black holes (SMBHs) as a novel probe for ultralight bosons. 
Our work consists of two parts. 
Firstly, in \cref{sec:QNMF_shift}, we demonstrate a new method to derive the effective potential of GW propagation due to a scalar field around a black hole from the scalar-field energy-momentum tensor. 
By substituting the effective potential into the equation governing gravitational perturbations, we can compute the shift of QNMFs due to the scalar field. 
We obtain the shift of QNMFs as a function of the mass of ultralight bosons and the cloud with logarithmic perturbation theory \cite{LPT}.
Secondly, in \cref{sec:Boson_search}, we explore the possibility of searching for ultralight bosons by measuring the QNMF shift of the ringdown phase of an extreme-mass-ratio inspiral (EMRI). 
We find that a single detection of the ringdown phase of an EMRI occurring at a nearby SMBH, such as Sagittarius A* (Sgr A*) and M32, enables us to rule out or confirm the existence of ultralight bosons of masses down to $\sim 10^{-17} \rm eV $. 
At last, in \cref{sec:conclusion}, we discuss the implications of our results. 

\section{Quasinormal-mode frequency shift due to bosonic clouds}
\label{sec:QNMF_shift}

\subsection{Assumptions and approximations} 

We consider a stellar-mass black hole of mass $ m $ spiraling into a host black hole of mass $ M $ with dimensionless spin $ a $, surrounded by a cloud of mass $ M_s $ formed by bosons of mass $\mu$, with $ m \ll M $ (an EMRI).
We assume that (\textbf{A1}) $ (M \mu)^2 \ll 1 $ and (\textbf{A2}) the smaller black hole does not disturb the bosonic cloud throughout the inspiral and ringdown phase, so the cloud can be described by the well-known bosonic wave function (see \cref{eq:wave_fn} below) during the ringdown phase.
(\textbf{A2}) is justified by numerical simulations of massive scalar hair around black holes \cite{baumann2019probing, berti2019ultralight}, where it was shown that the cloud depletes only a negligible fraction of its mass during an EMRI. 
Because the peak of the cloud's density goes as $(M\mu)^{-2}$ \cite{brito2015black}, by (\textbf{A1}) the black hole + cloud system's geometry is well described by the Kerr metric with the cloud treated as perturbation to the spacetime. 
Moreover, we will ignore frame-dragging when computing the cloud's effective potential, which introduces corrections of order $a (M \mu)^2 / M $ to the leading behavior of the cloud's effective potential.

\subsection{Effective potential of gravitational-wave propagation due to bosonic clouds}

We consider scalar ultralight bosons described by the Lagrangian density for a massive scalar field
\begin{equation}\label{eq:Lagrange_Density}
    \mathcal{L} = \frac{1}{2}\partial_\alpha \Phi \partial^\alpha \Phi + \frac{1}{2} \mu^2 \Phi^2,
\end{equation}
where $ \Phi$ is the wave function of the boson and $\mu$ is the mass of the boson. 
With this Lagrangian, one can obtain the Klein-Gordon equation governing the evolution of $\Phi$, 

\begin{equation}
\label{eq:KG_eqn_mu}
\Box \Phi (t, r, \theta, \phi) - \mu^2 \Phi (t, r, \theta, \phi) = 0, 
\end{equation}
where $ \Box = \frac{1}{\sqrt{-g}} \partial_{\mu} \left(\sqrt{-g} g^{\mu \nu} \partial_{\mu} \right) $ is the d'Alembertian operator. 
Since \cref{eq:KG_eqn_mu} is a separable partial differential equation, we consider \cite{ULB_freq_01, ULB_radial_function}
\begin{equation}
\Phi(t, r, \theta, t) = L_{n \ell m} (r) {S}_{\ell m} (\theta, \phi; \omega) e^{- i \omega_{n \ell m} t}, 
\end{equation}
where $L_{n \ell m}(r)$ is a function of $r$, $S_{\ell m}(\theta, \phi; \omega)$ is the spheroidal harmonic function of spin-weight 0, $n$ is the principal quantum number, $\ell$ is the azimuthal quantum number, and $m$ is the magnetic quantum number.
Assuming $(M \mu)^2 \ll 1 $ and solving \cref{eq:KG_eqn_mu} subjected to some physical boundary conditions \cite{detweiler1980klein}, one can obtain the characteristic oscillation frequency of the bosonic field \cite{Superradiance, detweiler1980klein, ULB_freq_01, ULB_freq_02, ULB_freq_03, Berti_review}
\begin{equation}\label{eq:mu_freq}
\begin{split}
\omega_{n \ell m} \sim & \mu-\frac{\mu}{2}\left(\frac{M \mu}{\ell+n+1}\right)^{2} \\
& +\frac{i}{\gamma_{n \ell m} M}\left(\frac{a m}{M}-2 \mu r_{+}\right)\left(M \mu\right)^{4 \ell+5}, 
\end{split}
\end{equation}
where $\gamma_{n \ell m}$ is a parameter which depends on $n$, $\ell$, and $m$. 

Inspecting \cref{eq:mu_freq}, one finds that the imaginary part of $\omega$ will be positive if $M \mu \sim 1 $ and $0 < \omega^{\rm Re}_{n \ell m} < m \Omega_{\rm H} $, where $ \Omega_{\rm H} = a / 2 r_+ $ is the angular velocity of the event horizon. 
This implies that $\Phi $ will be exponentially growing within this parameter space. 
This phenomenon is known as superradiant instability \cite{Superradiance, detweiler1980klein, ULB_freq_01, ULB_freq_02, ULB_freq_03}, through which ultralight bosons extract rotational energy of the host black hole to form a cloud. 
For the fastest-growing $n \ell m = $ 011 mode, the cloud is well described by the following wave function \footnote{
By using this function, it is implicitly approximated that $S_{11} \sim Y_{11}$, where $Y_{11}$ is the spherical harmonic for $\ell = m = 1 $. 
For our studies, where we assume $\mu \sim \omega $ to be small, it should be a good approximation \cite{spherodial_function}. }
 \cite{brito2015black}
\begin{equation}\label{eq:wave_fn}
\begin{split}
\Phi(t, r, \theta, \phi) = & \left( \frac{M_s}{\pi M}\right)^{\frac{1}{2}} \left( M \mu \right)^3 \mu r e^{-\frac{1}{2} M \mu^2 r} \\
& \times \sin \theta \cos \left (\phi - \omega^{\rm Re}_{011} t \right). 
\end{split}
\end{equation}
From \cref{eq:Lagrange_Density}, we can also derive the energy-momentum tensor of the bosonic cloud, 
\begin{equation}\label{eq:Stress-Tensor}
T_{\mu \nu} = -g_{\mu \nu} \left(\frac{1}{2}g^{\alpha \beta} \partial_\alpha \Phi \partial_\beta \Phi + \frac{1}{2} \mu^2 \Phi^2 \right) + \partial_{\mu} \Phi \partial_{\nu} \Phi.
\end{equation}

Gravitational perturbations around a rotating black hole are governed by the Teukolsky equation  \cite{teukolsky1973perturbations, Teukolsky_01_PRL, Teukolsky_03_ApJ, Teukolsky_04_ApJ},
and the effects of a bosonic cloud on GWs propagating around a black hole can be studied by solving the equation with the energy-momentum tensor of the cloud entering as the source term, 
\begin{equation}
\mathfrak{L} \psi = T_{\rm source}, 
\end{equation}
where $ \mathfrak{L} $ is a linear second-order partial differential operator \cite{teukolsky1973perturbations, Teukolsky_01_PRL, Teukolsky_03_ApJ, Teukolsky_04_ApJ},
\begin{equation}
\begin{aligned}
& \mathfrak{L} \\
& = \left[\frac{\left(r^{2}+ M^2 a^{2}\right)^{2}}{\Delta}-M^2 a^{2} \sin ^{2} \theta\right] \frac{\partial^{2} }{\partial t^{2}}+\frac{4 M^2 a r}{\Delta} \frac{\partial^{2} }{\partial t \partial \phi} \\
& +\left[\frac{M^2 a^{2}}{\Delta}-\frac{1}{\sin ^{2} \theta}\right] \frac{\partial^{2} }{\partial \phi^{2}} -\Delta^{-s} \frac{\partial}{\partial r}\left(\Delta^{s+1} \frac{\partial }{\partial r}\right) \\
& -\frac{1}{\sin \theta} \frac{\partial}{\partial \theta}\left(\sin \theta \frac{\partial }{\partial \theta}\right) -2 s\left[\frac{M a(r-M)}{\Delta}+\frac{i \cos \theta}{\sin ^{2} \theta}\right] \frac{\partial }{\partial \phi} \\
& -2 s\left[\frac{M\left(r^{2}-M^2 a^{2}\right)}{\Delta}-r-i M a \cos \theta\right] \frac{\partial }{\partial t} \\
& +\left(s^{2} \cot ^{2} \theta-s\right), 
\end{aligned}
\end{equation}
where $\Delta = (r - r_{+})(r - r_{-})$, $r_{+}$ and $r_{-}$ are, respectively, the outer and inner event horizon, $ \psi = \rho^{-4} \psi_4  $ is a perturbation function, $ \rho^{-1} = r - i M a \cos \theta $, $\psi_4$ is the fourth Weyl scalar, and $T_{\rm source}$ is the source term due to the bosonic cloud which acts as an external source that drives the gravitational perturbations. 
In the far-field limit, the gravitational perturbations are encoded as $ \psi_4 \propto \ddot{h}_{+} - i \ddot{h}_{\times}$. 
We assume separable solutions such that $ \psi_{\ell m} = R_{n \ell m}(r) S_{\ell m}(\theta, \phi) e^{-i \tilde{\omega}_{n \ell m} t }$, where $ S_{\ell m}(\theta, \phi)$ is the spheroidal function, $ R_{n \ell m} (r) $ is the radial part of the perturbation function and $ \tilde{\omega}_{n \ell m} $ is the QNMF of the $n \ell m$th mode \footnote{Not to be confused with the mode number of the bosonic cloud. $m$ is also not to be confused with $m^{\mu}$, one of the tetrad vectors. }. 
If a matter field such as a bosonic cloud surrounds the black hole, $ R_{n \ell m} (r) $ satisfies the radial master equation
\begin{equation}\label{eq:radial_TE_w_source}
\begin{split}
\Delta^{2} \frac{d}{d r}\left[\Delta^{-1} \frac{d R_{n \ell m}}{d r}\right]+V_{n \ell m}(r) R_{n \ell m}=\hat{T}_{n \ell m}. \\
V_{n \ell m}(r)=\left[\frac{K^{2}+4 i(r-M) K}{\Delta}+8 i \tilde{\omega}_{n \ell m} r+_{-2} \lambda_{n \ell m}\right], 
\end{split}
\end{equation}
where $V_{n \ell m} (r) $ is the effective potential for GW propagation, $K=(r^2+M^2 a^2)\tilde{\omega}_{n \ell m}-Mam$, $ -2 \lambda_{n \ell m}:=-2 A_{n \ell m}+M^2 a^{2} \tilde{\omega}_{n \ell m}^{2}-2 M a m \tilde{\omega}_{n \ell m}$ is the separation constant of the radial part \cite{Nakamura_01}, and $ A_{n \ell m} $ is the eigenvalue of the separated equation for $ S_{n \ell m }$. 
$ \hat{T}_{n \ell m} $ is the projected source term of the radial master equation, given by \cite{siemonsen2020gravitational}
\begin{equation}\label{eq:T_nlm}
\hat{T}_{n \ell m} = 4 \int_{\mathbb{R}} \frac{\mathrm{d}t}{\sqrt{2 \pi}}\int_{\mathbb{S}^2}\mathrm{d}\Omega S_{n \ell m}(\theta) e^{i \tilde{\omega}_{n \ell m} t - i m \phi} \frac{\Tilde{T}(t,r,\Omega)}{\rho^5 \rho^*}, 
\end{equation}
where 
\begin{equation}\label{eq:T}
\tilde{T}(t, r, \Omega)= \tilde{T}_{nn} + \tilde{T}_{nm} + \tilde{T}_{\bar{m} \bar{m}},
\end{equation}
and
\begin{equation}\label{eq:T_XX}
\begin{split}
\tilde{T}_{\bar{m} \bar{m}} =&-\frac{1}{4} \rho^{8} \bar{\rho} \Delta^{2} \hat{\mathcal{J}}_{+}\left[\rho^{-4} \hat{\mathcal{J}}_{+}\left(\rho^{-2} \bar{\rho} T_{\bar{m} \bar{m}}\right)\right], \\
\tilde{T}_{nn} =& -\frac{1}{2} \rho^{8} \bar{\rho} \hat{\mathcal{L}}_{-1}\left[\rho^{-4} \hat{\mathcal{L}}_{0}\left(\rho^{-2} \bar{\rho}^{-1} T_{n n}\right)\right] \\ 
\tilde{T}_{nm} =&-\frac{1}{2 \sqrt{2}} \rho^{8} \bar{\rho} \Delta^{2} \hat{\mathcal{L}}_{-1}\left[\rho^{-4} \bar{\rho}^{2} \hat{\mathcal{J}}_{+}\left(\rho^{-2} \bar{\rho}^{-2} \Delta^{-1} T_{\bar{m} n}\right)\right] \\ 
& -\frac{1}{2 \sqrt{2}} \rho^{8} \bar{\rho} \Delta^{2} \hat{\mathcal{J}}_{+}\left[\rho^{-4} \bar{\rho}^{2} \Delta^{-1} \hat{\mathcal{L}}_{-1}\left(\rho^{-2} \bar{\rho}^{-2} T_{\bar{m} n}\right)\right],\\
T_{n n} & :=T_{\mu \nu} n^{\mu} n^{\nu}, \\
T_{\bar{m} n} & :=T_{\mu \nu} n^{\mu} \bar{m}^{\nu}, \\
T_{\bar{m} \bar{m}} & :=T_{\mu \nu} \bar{m}^{\mu} \bar{m}^{\nu}, \\
n^{\mu} &:=\frac{1}{2 \Sigma}\left(r^{2}+M^2 a^{2},-\Delta, 0, Ma\right), \\ 
m^{\mu} &:=\frac{\rho^{*}}{\sqrt{2}}\left(i M a \sin \theta, 0,1, i \csc \theta\right), \\
\Sigma & = r^2 + M^2 a^2 \cos^2 \theta, \\ 
\hat{\mathcal{L}}_{s} & :=\partial_{\theta}-i \csc \theta \partial_{\phi}-i a \sin \theta \partial_{t}+s \cot \theta. \\ 
\hat{\mathcal{J}}_{+} & :=\partial_{r}-\Delta^{-1}\left[\left(r^{2}+M^2 a^{2}\right) \partial_{t} + M a \partial_{\phi}\right]. 
\end{split}
\end{equation}
At spatial infinity, $ m^{\mu}$ approaches the limit
\begin{equation}
\begin{aligned}\label{eq:Tetrad}
m^{\mu} &\rightarrow \frac{1}{\sqrt{2}} (\hat{\theta}^{\mu} + i \hat{\phi}^{\mu}). 
\end{aligned}
\end{equation}

As the host black hole is subject to gravitational perturbations, due to both the capture of a less massive black hole as well as the monochromatic GWs generated by the bosonic cloud, we need to replace $ g_{\mu \nu} $ in \cref{eq:Stress-Tensor} by
\begin{equation}\label{eq:metric_inclusions}
\begin{split}
g_{\mu \nu} & = g_{\mu \nu}^{(0)} + h_{\mu \nu}, \\
h_{\mu \nu} & = h^{\rm (RD)}_{\mu \nu} + h^{\rm (ULB)}_{\mu \nu},
\end{split}
\end{equation}
where $ g_{\mu \nu}^{(0)} $ is the Kerr metric describing the spacetime around the host black hole, $ h^{\rm (ULB)}_{\mu \nu}$ represents GWs emitted by the boson could, and $h^{\rm (RD)}_{\mu \nu}$ represents GWs during the ringdown phase due to the black hole merger. 
Then we can split $ \tilde{T}(t, r, \Omega)$ into three parts, 
\begin{equation}
\begin{split}
\tilde{T}(t, r, \Omega) & = \tilde{T}^{(\rm ULB)}_{1} + \tilde{T}^{(\rm ULB)}_{2} + \tilde{T}^{(\rm RD)} , \\
\tilde{T}^{(\rm ULB)}_{1} & = \tilde{T}^{(\rm ULB)}_{1}[g_{\mu \nu}^{(0)}\partial_\alpha \Phi \partial^\alpha \Phi, \Phi  g_{\mu \nu}^{(0)} \Phi] \\
\tilde{T}^{(\rm ULB)}_{2} & = \tilde{T}^{(\rm ULB)}_{2}[h^{\rm (ULB)}_{\mu \nu} \partial_\alpha \Phi \partial^\alpha \Phi, \Phi  h^{\rm (ULB)}_{\mu \nu} \Phi] \\
\tilde{T}^{(\rm RD)} & = \tilde{T}^{(\rm RD)}[h^{\rm (RD)}_{\mu \nu} \partial_\alpha \Phi \partial^\alpha \Phi, \Phi  h^{\rm (RD)}_{\mu \nu} \Phi], 
\end{split}
\end{equation}
where $ \tilde{T}^{(\rm ULB)}_{1}[g_{\mu \nu}^{(0)}\partial_\alpha \Phi \partial^\alpha \Phi, \Phi  g_{\mu \nu}^{(0)} \Phi]  $ is obtained by projecting the terms in $ T_{\mu \nu }$ containing $ g_{\mu \nu}^{(0)}\partial_\alpha \Phi \partial^\alpha \Phi$ and $  \Phi  g_{\mu \nu}^{(0)} \Phi $ onto $ n^{\nu} $ and $ m^{\nu}$ according to \cref{eq:T_XX}, and similarly for $\tilde{T}^{(\rm ULB)}_{2} $ and $ \tilde{T}^{(\rm RD)} $.

Each of the above terms has its own effects on GW generation. 
To see this, we split $ \psi $ into three parts 
\begin{equation}
\psi = \psi^{(\rm ULB)}_{1} + \psi^{(\rm ULB)}_{2} + \psi^{(\rm RD)}, 
\end{equation}
and each of these terms satisfy their own Teukolsky equation with the corresponding source term. 
For $ \psi^{(\rm ULB)}_{1} $, we have 
\begin{equation}
\mathfrak{L} \psi^{(\rm ULB)}_{1} = \tilde{T}^{(\rm ULB)}_{1}[g_{\mu \nu}^{(0)}\partial_\alpha \Phi \partial^\alpha \Phi, \Phi  g_{\mu \nu}^{(0)} \Phi]. 
\end{equation}
This equation represents the quasimonochromatic emission of GWs of frequency $ \sim 2 \mu $ by the bosonic cloud, which slowly dissipates the cloud on timescales significantly greater than that of a ringdown detection.
Coherent searches could identify these waves in the stochastic GW background (see, e.g., \cite{CW_01}). 
The next term leading to quasimonochromatic GWs is given by,
\begin{equation}
\mathfrak{L} \psi^{(\rm ULB)}_{2} = \tilde{T}^{(\rm ULB)}_{2}[h^{\rm (ULB)}_{\mu \nu} \partial_\alpha \Phi \partial^\alpha \Phi, \Phi  h^{\rm (ULB)}_{\mu \nu} \Phi]. 
\end{equation}
This represents the generation of the next-leading order long-lived GWs, of the frequency of $ \sim 4 \mu, $ by the bosonic cloud. 
Similarly, these waves could be detected using coherent analysis. 
Finally, we have
\begin{equation}\label{eq:TE_ULB_operator}
\mathfrak{L} \psi^{(\rm RD)} = \tilde{T}^{(\rm RD)}[h^{\rm (RD)}_{\mu \nu} \partial_\alpha \Phi \partial^\alpha \Phi, \Phi  h^{\rm (RD)}_{\mu \nu} \Phi]. 
\end{equation}
This equation does not depend on $ \psi^{(\rm ULB)}_{1} $, $ \psi^{(\rm ULB)}_{2} $ and $ h^{\rm (ULB)}_{\mu \nu} $. 
Therefore, this is an equation decoupled from the previous two.
\cref{eq:TE_ULB_operator} shows that metric perturbations around the black hole, corresponding to ringdown GWs, interact with the surrounding bosonic cloud.
In this work, we focus on the effects by $ h^{\rm (RD)}_{\mu \nu} $ as it can modify the ringdown waveform and can be measured independently from the continuous emission $ h^{\rm (ULB)}_{\mu \nu} $. 

Focusing on \cref{eq:T_XX} and \cref{eq:TE_ULB_operator}, we find that $ T_{\bar{m} \bar{m}} $ in the far field can be expressed in terms of $ \psi_4 $.
We start with
\begin{equation}
\begin{split}
T_{\bar{m} \bar{m}} & = \bar{m}^{\beta} \bar{m}^{\gamma}T_{\beta \gamma}  \\
& = \frac{1}{2} \bar{m}^{\beta} \bar{m}^{\gamma} \left[ - g_{\beta \gamma} (\partial_{\alpha} \Phi \partial^{\alpha} \Phi + \mu^2 \Phi^2 ) + \partial_{\beta} \Phi \partial_{\gamma} \Phi \right] \\
& = \frac{1}{2} \bar{m}^{\beta} \bar{m}^{\gamma} \left[ - (g^{(0)}_{\beta \gamma}+h_{\beta \gamma}) (\partial_{\alpha} \Phi \partial^{\alpha} \Phi + \mu^2 \Phi^2 )  \right] + (\bar{\delta} \Phi)^2, 
\end{split}
\end{equation}
where $h_{\beta \gamma} = h^{\rm (RD)}_{\beta \gamma}$ (from now on we suppress the superscript ``(RD)") and $\bar{\delta} \Phi = \bar{m}^\beta \partial_\beta \Phi$. 
By the null-cross normalization condition, we have $ g^{(0)}_{\alpha \beta} \bar{m}^{\alpha} \bar{m}^{\beta} = 0$ \cite{NP_original_paper, Schnittman_2008, Misner:1973prb}, thus 
\begin{equation}
T_{\bar{m} \bar{m}} = - \frac{1}{2} \bar{m}^{\beta} \bar{m}^{\gamma} \left[ h_{\beta \gamma} (\partial_{\alpha} \Phi \partial^{\alpha} \Phi + \mu^2 \Phi^2 )  \right] + (\bar{\delta} \Phi)^2. 
\end{equation}
Since single-mode ringdown waves can be expressed in terms of damped sinusoidal waves, the perturbation $ h_{\beta \gamma} $ of a particular quasinormal mode can be written as 
\begin{equation}
h_{\beta \gamma} = A_{\beta \gamma}(r, \theta, \phi) e^{-i \omega t}, 
\end{equation}
where $A_{\beta \gamma} $ is a tensorial function of position and $ \omega $ is a complex frequency. 
Hence, we can write 
\begin{equation}
h_{\beta \gamma} \approx - \frac{1}{\omega^2} \ddot{h}_{\beta \gamma}.   
\end{equation}
By (\textbf{A1}), the most important region for gravitational interaction between ringdown waves and the bosonic cloud is in the far field. 
The far-field limit of the definition of the fourth Weyl scalar gives \cite{Schnittman_2008, Misner:1973prb}
\begin{equation}
\begin{split}
\psi_4 (r \rightarrow + \infty) & = \frac{1}{2} \left( \ddot{h}_{+} - i \ddot{h}_{\times} \right) \\
& = \frac{1}{4} \left( \ddot{h}_{\theta \theta} - \ddot{h}_{\phi \phi} \right) - \frac{1}{2} i \ddot{h}_{\theta \phi} \\
& = \frac{1}{2}\bar{m}^{\beta} \bar{m}^{\gamma} \ddot{h}_{\beta \gamma} \\
& = - \frac{1}{2}\omega^2 \bar{m}^{\beta} \bar{m}^{\gamma} h_{\beta \gamma}.
\end{split}
\end{equation}
As $ (\bar{\delta} \Phi)^2 $ has no explicit dependence on $ \psi_4 $, it will not contribute as an additional effective potential, but only as an extra excitation of quasinormal modes,
Thus, we can suppress it from $  T_{\bar{m}\bar{m}} $ if we only want to calculate the shifts in the QNMFs. 
Focusing on $T_{\bar{m}\bar{m}}$, we have
\begin{equation}
\begin{split}
    T_{\bar{m}\bar{m}} & \approx \frac{1}{\omega^2 } \psi_4 [\partial_\alpha \Phi \partial^\alpha \Phi +\mu^2 \Phi^2] \\
    & = \frac{1}{\omega^2} \rho^4 \psi [\partial_\alpha \Phi \partial^\alpha \Phi +\mu^2 \Phi^2],
\end{split}
\end{equation}
indicating that $T_{\bar{m}\bar{m}}$ carries a factor of the GW waveform $\psi$ and can thus be considered as a term contributing to the effective potential of wave propagation.
We note that $T_{nm}$ and $T_{nn}$ contain no such factors of $\psi_4$, so they neither affect the effective potential of wave propagation nor shift the QNMFs.
Effectively, we consider 
\begin{equation}\label{eq:T_and_psi}
\begin{split}
&\tilde{T} (t, r, \theta, \phi) \\
&= - \frac{1}{4 \omega^2} \rho^{8} \bar{\rho} \Delta^{2} \hat{\mathcal{J}}_{+}\left\{\rho^{-4} \hat{\mathcal{J}}_{+}\left[\rho^{2} \bar{\rho} \left( \partial_\alpha \Phi \partial^\alpha \Phi +\mu^2 \Phi^2 \right) \rho^4 \psi \right] \right\}, 
\end{split}
\end{equation}
where terms that do not contain $\psi_4 $ have been suppressed. 

We then calculate $\hat{T}_{n \ell m}$ of \cref{eq:radial_TE_w_source} from \cref{eq:T_and_psi} using \cref{eq:T_nlm} and then expressing $\hat{T}_{n \ell m}$ in terms of $ R_{n \ell m}(r) $,
\begin{equation}
    \hat{T}_{n \ell m} = R_{n \ell m}(r) V^{(\rm ULB)}_{n \ell m}(r).
\end{equation}
As the ringdown phase of an EMRI is dominated by the $n \ell m = 022 $ and $021$ modes (see later discussion), we focus on calculating the shifts of these two modes. 
We choose to use two modes instead of one to reduce potential degeneracy introduced by the addition of the $M\mu$ parameter during Bayesian inference in \cref{sec:Boson_search}.
Note that these modes are not to be confused with the dominant excitation mode of the bosonic cloud.

By (\textbf{A1}), the gravitational interaction between the cloud and GWs should be the most important when $ r \gg r_+ $ where rotational frame dragging is negligible. 
Thus, we can set $a = 0$ to reflect negligible frame dragging in the far field \footnote{Note that such $a = 0$ assumption is only valid when we are considering the interaction of GWs with the cloud when they meet the cloud potential barrier in the far field.
When calculating the shape of the cloud and the usual potential barrier, we still have to include a nonzero spin, or else the cloud will not form in the first place}.
Then we have, for a boson cloud of $n = 0$, $\ell = m = 1 $, and ringdown waves of $n = 0, \ell = 2$, and $m = 1, 2$, 
\begin{align}
\begin{split}\label{eq:VULB}
    V^{\rm (ULB)}_{0 2 m}(r) = & \frac{(r^2+M^2a^2)^2}{\Delta}\frac{4 M_s \tilde{\mu }^8 e^{-\tilde{r}\tilde{\mu }^2}}{7 \pi  M^3 \tilde{k}^2 \tilde{r}^6} \sum_{n = 0}^{2} \tilde{\mu}^n C_n^{0 2 m} (\tilde{r}, \tilde{k}), 
\end{split}
\end{align}
where $\tilde{k} = i M \omega$, $\tilde{\mu} = M \mu$, and $\tilde{r} = r/M$.
$ C_n^{0 2 m}$ with $m = 1, 2$ are polynomial functions with explicit forms given in \cref{sec:App_A}. 
Thus \cref{eq:radial_TE_w_source} becomes 
\begin{equation}\label{eq:radial_TE_modified}
\begin{split}
\Delta^{2} \frac{d}{d r}\left[\Delta^{-1} \frac{d R_{n \ell m}}{d r}\right]+\left[ V_{n \ell m}(r) + V^{\rm (ULB)}_{n \ell m}(r) \right] R_{n \ell m} = 0. 
\end{split}
\end{equation}

\begin{figure}[tp!]
\includegraphics[width=\columnwidth]{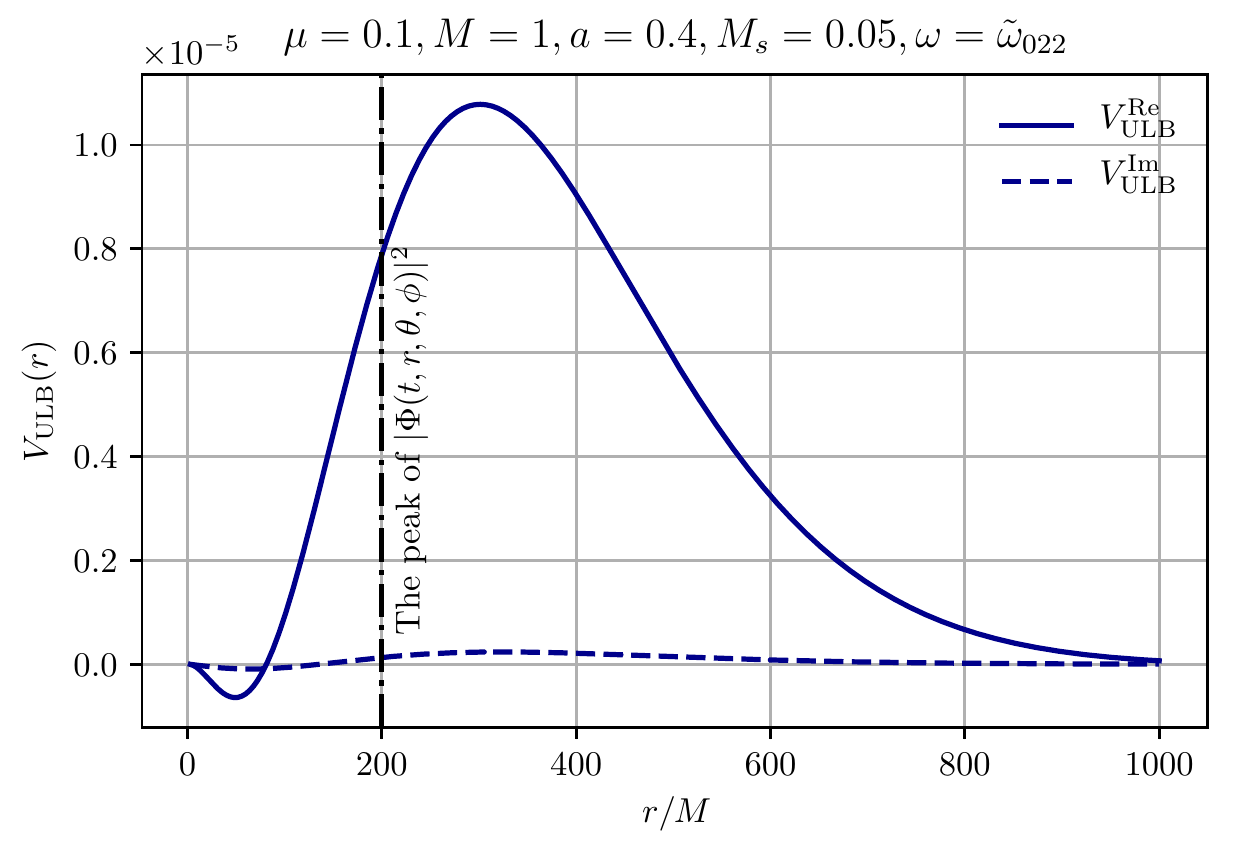}
 \caption{The real and imaginary parts of the effective potential for $n \ell m = 022$ mode GW propagation due to an ultralight-bosonic cloud around a black hole. 
We assume that the mass of the cloud is  $ M_s = 0.05 $, formed by bosons of mass $ \mu = 0.1 $ around a black hole of mass $ M = 1$ and $a=0.4$.
We assume that the GWs are propagating at the (complex) frequency of the 022 quasinormal mode.
The peak of the real potential is located close to that of $ |\Phi|^2 $, where $ \Phi $ is the wave function of ultralight bosons (at a given time and angular position), much further away from the event horizon.}
 \label{fig:V_ULB}
\end{figure}

\cref{fig:V_ULB} plots the effective potential $ V^{(\rm ULB)}_{022}(r) $ for the GW propagation built up by a bosonic cloud of $ M_s = 0.05 $, $ \mu = 0.1 / M = 0.1 $  around a black hole of $ M = 1 $, $a=0.4$.
This value of $a$ is close to that of the measured spin of Sagittarius A* \cite{Sgr_A_spin}.
This potential barrier affects the $ \omega = \tilde{\omega}_{022}$ mode, the dominant quasinormal mode during the ringdown phase of an EMRI \cite{hadar2011comparing}. 
The potential barrier corresponding to the 021 mode is qualitatively similar to that of the 022 mode. 
We find that $ V^{\rm Re}_{\rm ULB} $ peaks at far from the event horizon, close to the peak of $ |\Phi(t, r, \theta, \phi )|^2 $ (at a given time and angular position) which defines the mass density of the cloud. 
This similarity makes sense if we interpret $ V_{n \ell m}^{(\rm ULB)} $ as representing the effects of the cloud's gravity on the ringdown GWs. 
For higher spin, the effective potentials due to a bosonic cloud are qualitatively similar. 

\subsection{Shift of quasinormal-mode frequencies}

$ V^{(\rm ULB)}_{n \ell m}(r) $ and $ V_{n \ell m}(r) $ together form a new effective potential of GW propagation, selecting GWs of a different set of QNMFs to reach spatial infinity as purely outgoing waves. 
For simplicity, our work focuses on measuring or constraining $ \mu $ solely by measuring the shift of QNMFs of an SMBH surrounded by a bosonic cloud.
Since $ (M \mu)^2 \sim 10^{-2} \ll 1 $, $|V_{n \ell m}^{(\rm ULB)}| \sim (M \mu)^8 \ll |V_{n \ell m}|$, and we can solve \cref{eq:radial_TE_modified}  for $ \tilde{\omega}_{n \ell m}$ using logarithmic perturbation theory \cite{LPT}. 
Up to leading order, QNMFs in the source frame are given by
\begin{equation}\label{eq:QNM_LPT}
\tilde{\omega}_{n \ell m} \approx \tilde{\omega}_{n \ell m}^{(0)} + \frac{\Delta_{n \ell m}(\mu, M_s, M)}{2 \tilde{\omega}_{n \ell m}^{(0)}}, 
\end{equation}

\begin{figure}[tp!]
 \includegraphics[width=\columnwidth]{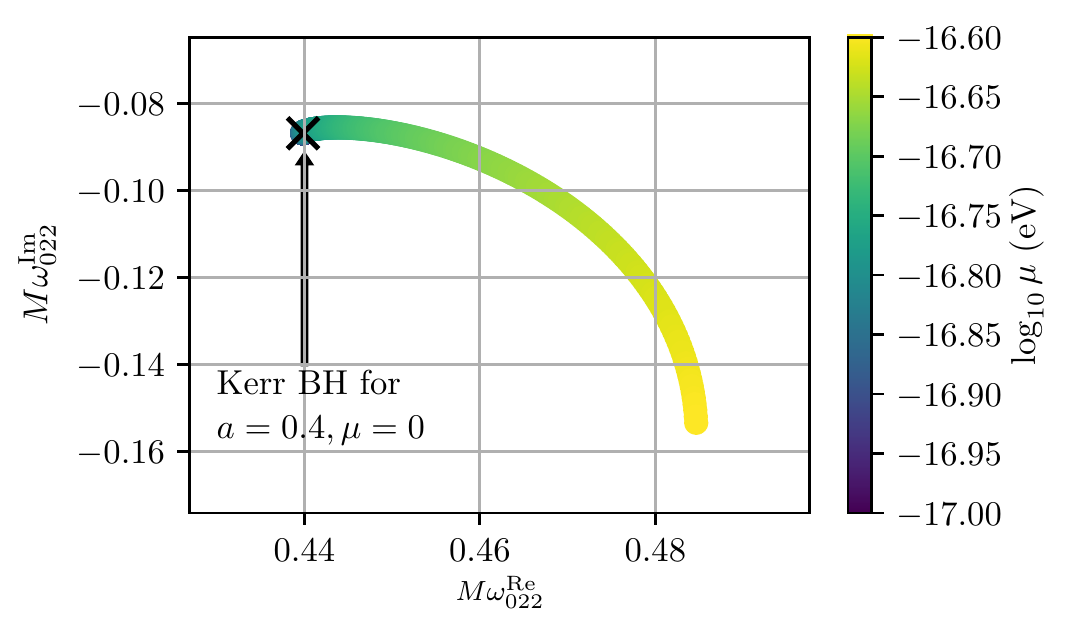}
 \caption{The shift of the 022 mode QNMFs for different $ \mu $.
 We assume $ M = 4.4 \times 10^6 M_{\odot}$ and dimensionless spin $ a = 0.4$ surrounded by a cloud of $\alpha = 0.05 $. 
 The cross marks the quasinormal-mode frequency of a Kerr black hole with $a = 0.4 $ for $\mu = 0 $. 
 As $ \mu $ increases, the 022-mode real frequency increases but the lifetime decreases, which is consistent with the changes of QNMF due to a dark matter halo of constant density.
 For $\mu$ smaller than $10^{-16.8}\rm eV$, the QNMF shift is masked by the green hue and the cross. 
 }
 \label{fig:M_Omega_022}
\end{figure}
for the 021 and 022 modes of ringdown waves, 
\begin{align}
\begin{split}\label{eq:Delta}
    \Delta_{0 2 m} = \, &\frac{32 M_s \tilde{k}^3 \tilde{\mu }^8 e^{-2 \left(2 \tilde{k}+\tilde{\mu }^2\right)}}{F(\tilde{k})}\\
    \times &\left[\frac{e^{-4 \tilde{k}}}{\tilde{b}^7}\left(\sum_{n = 0}^{16} \tilde{\mu}^n D_n^{0 2 m}(\tilde{k}) \right) - 4 e^{2 \tilde{\mu }^2} E^{0 2 m}(\tilde{k}, \tilde{\mu}, \tilde{b}) \right]\\
   F(\tilde{k}) = &64 \tilde{k}^4-56 \tilde{k}^3+36 \tilde{k}^2-15 \tilde{k}+3,\\
\end{split}
\end{align}
where $\tilde{k} = i M \omega$, $\tilde{\mu} = M \mu$, and $\tilde{b} = (2\tilde{k} - \tilde{\mu}^2)$.
The explicit expressions of polynomials $D_n^{0 2 m}(x)$ and functions $E^{0 2 m}(x)$ for $m = 1, 2$ are given in \cref{sec:App_A}. 

Fig.~\ref{fig:M_Omega_022} plots the trajectory on the complex plane of $ M \tilde{\omega}_{022}$ with $a = 0.4$ as $ \mu $ increases from $ 10^{-17} $ to $ 10^{-16.6} \rm eV $ (such that $M \mu \sim 1$), assuming $ \alpha = 0.05$. 
We assume $a=0.4$ because it is close to the spin of the Sagittarius A*,the closest supermassive black hole. 
For the ease of comparison with the case of $\mu = 0 $, the cross marks the quasinormal-mode frequency of a Kerr black hole of $a = 0.4 $ for $\mu = 0 $ (taken from \cite{Berti_QNMFs}). 
A yellower hue represents a larger $\mu$ and a bluer hue represents a smaller $\mu$. 
As $ \mu $ increases, both the real part and imaginary part of $\tilde{\omega}_{022}$ decreases; thus, the life time, which is consistent with the quasinormal mode shifts for a black hole surrounded by a constant density dark matter halo \cite{barausse2014can}.
For $\mu$ smaller than $\sim 10^{-16.8}\rm eV$, the QNMF shift is masked by the green hue and the cross. 
For higher spin such as $a = 0.9$, the 022-mode frequency follows a quantitatively similar trajectory starting from $M \omega = 0.672 - 0.065i$. 

\section{Implications on astrophysical observations}
\label{sec:Boson_search}

As the boson mass changes the gravitational QNMFs, it can be measured or constrained by observing the ringdown of astrophysical black holes. 
Possible sources for which our ringdown analysis could be applied to probe ultralight bosons include intermediate-mass ratio inspiral events or EMRIs, as cloud depletion in these cases are negligible \cite{bertone2019gravitational}.
We perform a mock data challenge to estimate our ability of measuring ultralight boson mass upon detecting the ringdown phase of an EMRI from Sgr A* and M32 by the Laser Interferometer Space Antenna (LISA), a proposed space-based interferometer capable of detecting high-mass mergers \cite{LISA_proposal, babak2017science, robson2019construction, Berti_QNM, SMBH_mass_fn_01, LISA_sensitivity_volume, SMBH_mass_fn_01_2, SMBH_mass_fn_03, EMRI_rate_01}. 

\subsection{Constructing the likelihood}

We construct a waveform model to simulate GWs during the ringdown phase. 
In the time domain, GWs emitted during the ringdown phase due to an EMRI can be expressed as a linear combination of damped sinusoids \cite{EMI_RD_Amp, LIGO_15}, 
\begin{equation}\label{eq:TD_WF}
h(t) = \frac{m}{r} \sum_{\mathit{n \ell m}} A_\mathit{n \ell m} e^{-i\phi_{n \ell m}-i \tilde{\omega}_\mathit{n \ell m} t}. 
\end{equation}
Here $m$ is the mass of the compact object inspiraling into the central host black hole and $r$ is the luminosity distance to the host black hole. 
$ A_\mathit{n \ell m} $ is the mode amplitude, and we use $ A_{022} \approx 10^{-0.1} $ and $A_{021} \approx 10^{-0.4} $, which are typical values for EMRIs \cite{EMI_RD_Amp}.
$ \phi_\mathit{n \ell m }$ is the initial phase of the $ n \ell m $th mode. 
The QNMFs $ \tilde{\omega}_{n \ell m} $ are functions of $ \mu $, $ M_s $, $ M $, and $ a $. 

When estimating the parameters, it will be more convenient to work in the frequency domain. 
We perform a Fourier transform on \cref{eq:TD_WF} following the \textit{FH convention} \cite{Berti_QNM}. 
The convention replaces $ e^{-i \omega_{n \ell m} t - t / \tau_{n \ell m}} $ by $ e^{-i \omega_{n \ell m} t - |t| / \tau_{n \ell m}} $ assuming that the ringdown starts at $ t = 0 $, 
\begin{equation}
\begin{split}
& \int_{-\infty}^{\infty} e^{i \omega t}\left(e^{- i \omega_{n \ell m} t-|t| / \tau_{n \ell m}}\right) d t\\
& =\frac{2 / \tau_{n \ell m}}{\left(1 / \tau_{n \ell m}\right)^{2}+\left(\omega - \omega_{n \ell m}\right)^{2}}. 
\end{split}
\end{equation}
Noting that $ \omega^{\rm Im}_{n \ell m} = -\tau^{-1}_{n \ell m} $ because the imaginary part of QNMFs are always negative, we arrive at  \footnote{In principle, our waveform model should also depend on the cosmological redshift and the sky position of the host SMBH. 
However, as we are considering Sgr A* and M32, whose redshift $z \sim 0 $, we omit $z$ from our inference. 
To have fair comparison with other proposed searches for dark matter with LISA such as \cite{hannuksela2019probing, DM_spike_01}, we assume that the detectors are oriented optimally for the plus polarized waves.
}
\begin{equation}\label{eq:FD_WF}
\begin{split}
& h(f; M_f, a_f, \mu, M_s, A_{n \ell m}, \phi_{n \ell m }) \\
& = - \frac{2m}{r} \sum_{n \ell m} \frac{A_{n \ell m} \omega_{\text{Im}, n \ell m} e^{- i \phi_{n \ell m}}}{\omega^{2}_{\text{Im}, n \ell m} + (2\pi f - \omega^{2}_{\text{Re}, n \ell m})^{2}}. 
\end{split}
\end{equation}

Upon detecting the GW strain data $\tilde{d}$ of an EMRI, we can estimate $M$, $a$, $ \mu $, $M_s$, $A_{n \ell m }$, and initial phase $\phi_{n \ell m}$ using Bayesian inference \footnote{Following other works which propose searches for ultralight bosons (see e.g. \cite{eda2013new}, \cite{hannuksela2019probing}), we assume that no other type of matter surrounds the black hole. }.
By Bayes' theorem, the posterior of the parameters describing the host black hole ($\vec{\theta}$) and bosonic cloud ($\alpha, \mu$) is 
\begin{equation}
p(\vec{\theta}, \alpha, \mu|\tilde{d}, H, I) \propto p(\tilde{d}|\vec{\theta}, \alpha, \mu, H, I) p(\vec{\theta}, \alpha, \mu, H, I), 
\end{equation}
where $p(\vec{\theta}, \alpha, \mu, H, I)$ is the prior of $\vec{\theta}, \alpha$, and $\mu$ (see \cref{table:prior} for the complete list of priors), while  $p(\tilde{d}|\vec{\theta}, \alpha, \mu, H, I)$ is the likelihood
\begin{equation} \label{eq:loglikelihood}
\mathcal{L}(s|\vec{\theta}) \propto \exp \left( - \frac{1}{2} \braket{s-h(\vec{\theta})|s-h(\vec{\theta})}\right).
\end{equation}
$\tilde{d} = \tilde{n} + \tilde{h}(\vec{\theta}_{\rm inj}, \alpha_{\rm inj}, \mu_{\rm inj}) $ is the injected strain data, including noises, $\tilde{n}$, simulated according to the power spectral density (PSD) of LISA \cite{robson2019construction} and the injected signal $\tilde{h}(\vec{\theta}_{\rm inj}, \alpha_{\rm inj}, \mu_{\rm inj}) $ with $\vec{\theta}_{\rm inj}, \alpha_{\rm inj}, \mu_{\rm inj}$ being the injected parameters.
The inner product $\braket{a|b}$ is defined as
\begin{equation}
 \braket{a|b} = 4 \Re \left[ \int_{f_{\rm low}}^{f_{\rm high}} \frac{a(f) b^*(f)}{S_n(f)} df \right], 
\end{equation}
where $f_{\rm low} = 10^{-4} \, \rm Hz $ and $f_{\rm high} = 1 \, \rm Hz $ are the lower and upper limit of LISA's sensitivity band, respectively, while $S_n(f)$ is the PSD of LISA's noise. 

\begin{table}[tp!]
\centering
\begin{tabular}{llll}
\toprule
Variables  &   Prior type  &  Range \\ \hline
$M_f$  &  Log uniform  &   $[10^{6}, 10^{8}] M_{\odot}$ \\
$a_f$  &  Uniform  &   $[0, 1]$ \\
$\alpha$  &  Uniform  &  $[0, 0.2]$ \\
$\mu$  &  Conditional and log uniform  &   $[10^{-18}, M_f^{-1}] \rm eV $ \\
$A_{n \ell m}$  &  Log uniform  &   $[10^{-22}, 10^{-13}] $ \\
$\phi_{n \ell m}$  &  Uniform  &   $[0, 2 \pi ] $ \\
\toprule
\end{tabular}
\caption{The prior prescribed for various parameters for the mock data challenge.}
\label{table:prior}
\end{table}

Specifically, our mock data simulate the ringdown phase of EMRIs due to compact objects plunging into Sgr A* and M32.
The mass of the compact object is set to be $10 M_{\odot}$, typical of stellar mass black holes inspiraling into a supermassive black hole \cite{EMRI_rate_01, EMRI_rate_02}. 
For Sgr A*, we take $a = 0.4 $, which is its measured spin \cite{Sgr_A_spin}.
As for M32, since its host black-hole spin is not well measured, we assume $a \sim 0.9$, a common spin of SMBHs \cite{Reynolds_2013}.
The mode component $ A_{n \ell m} $ of the injected signal is computed according to \cite{hadar2011comparing}. 
We include only the 022 and 021 modes in mock signals as they are the dominant quasinormal modes of the EMRI ringdown signals  \cite{hadar2011comparing}. 
We estimated that the detectable EMRI ringdown phase of Sgr A* and M32 can last for $\sim 10 $ times the lifetime of the 022 mode. 
We found that the optimal signal-to-noise ratio (SNR, $\rho_{\rm opt}$), defined by 
\begin{equation}
\rho_{\rm opt}^2 = \braket{\tilde{h}|\tilde{h}} = 4 \Re \left[ \int_{f_{\rm low}}^{f_{\rm high}} \frac{\tilde{h}(f) \tilde{h}^{*}(f)}{S_n(f)} df \right], 
\end{equation}
of the EMRI ringdown of Sgr A* is $\sim 1.38 \times 10^{5} $ and that of M32 is $\sim 1.6 \times 10^3 $.

\subsection{Results of mock data challenge}

\begin{figure}[tp!]
 \includegraphics[width=\columnwidth]{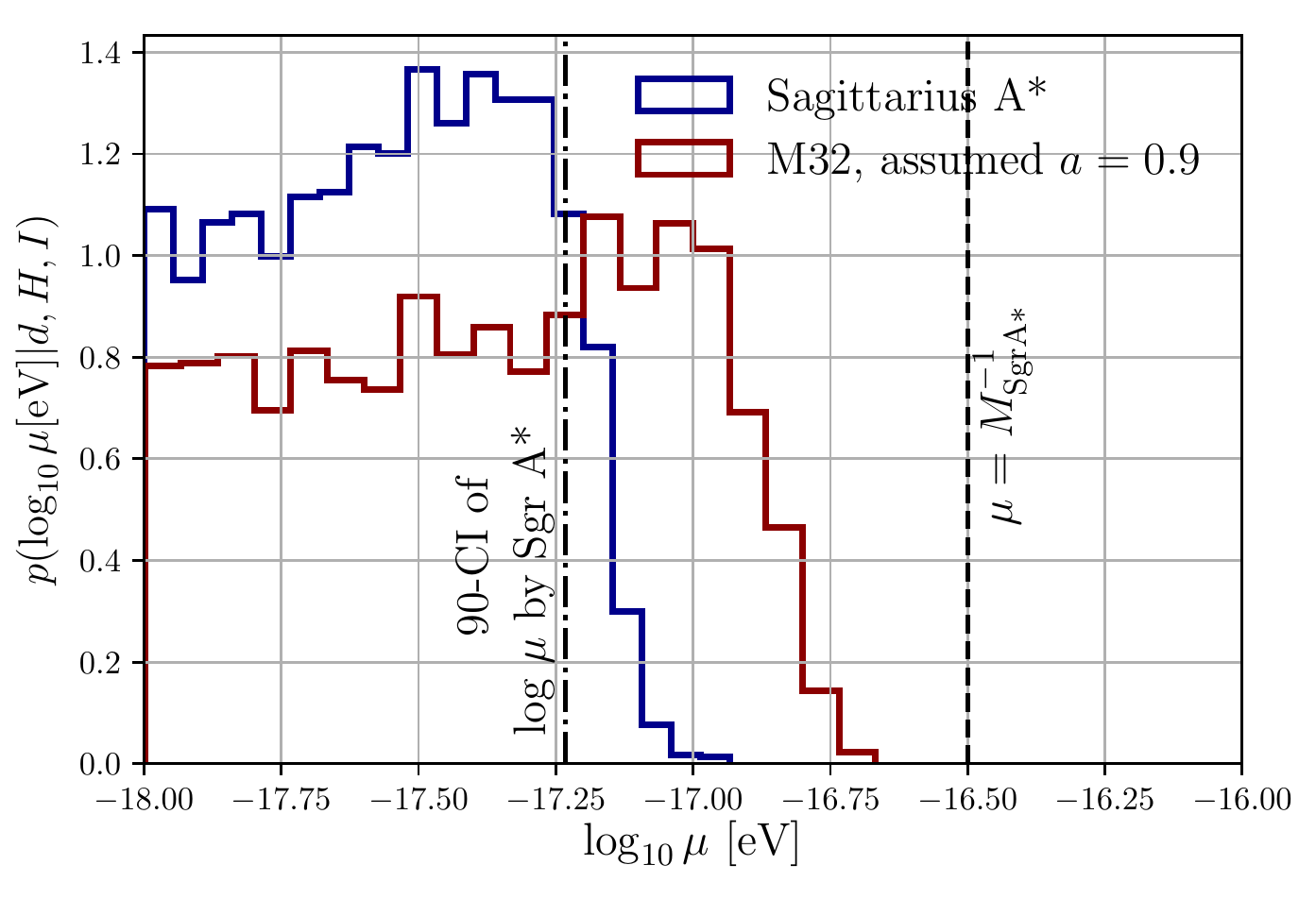}
 \caption{Constraints on $ \mu $ by a LISA detection of the EMRI ringdown phase of Sagittarius A* ($ = 4.3 \times 10^6 M_{\odot}$ \cite{Sgr_A_mass}, in blue) and M32 ( $ = 3.4 \times 10^6 M_{\odot}$ \cite{M32_mass}, in red). 
 We assume $\mu = 0 $ and carry out our ringdown analysis over ten lifetimes of the 022 modes, with the mass, spin and luminosity distance set to be close to that of the corresponding SMBH. 
 By observing the ringdown phase of an EMRI of these SMBHs, we can rule out $ \mu \sim 10^{-17} \rm eV $, as shown by the 90\% confidence interval of our posterior for Sagittarius A* marked by the dashed-dotted vertical line. 
 }
\label{fig:PDF_mu}
\end{figure}

Fig.~\ref{fig:PDF_mu} shows the marginalized posterior of the base-10 log of boson mass $ \mu $ in eV inferred from the mock ringdown signal of Sgr A* (solid blue line) and M32 (solid red line) with an injected null-hypothesis signal $\mu = 0$. 
The posterior of $ \log_{10} \mu $ resembles a step-function shape with a steep cutoff at $ \sim 10^{-17} \rm eV$. 
Beyond the cutoff, the posterior distributions show no support, thereby putting a bound on $ \mu $. 
The posteriors demonstrate that such an analysis on a single ringdown event can lead to stringent constraints on boson mass if no QNMF shift is measured.

\begin{figure}[tp!]
 \includegraphics[width=\columnwidth]{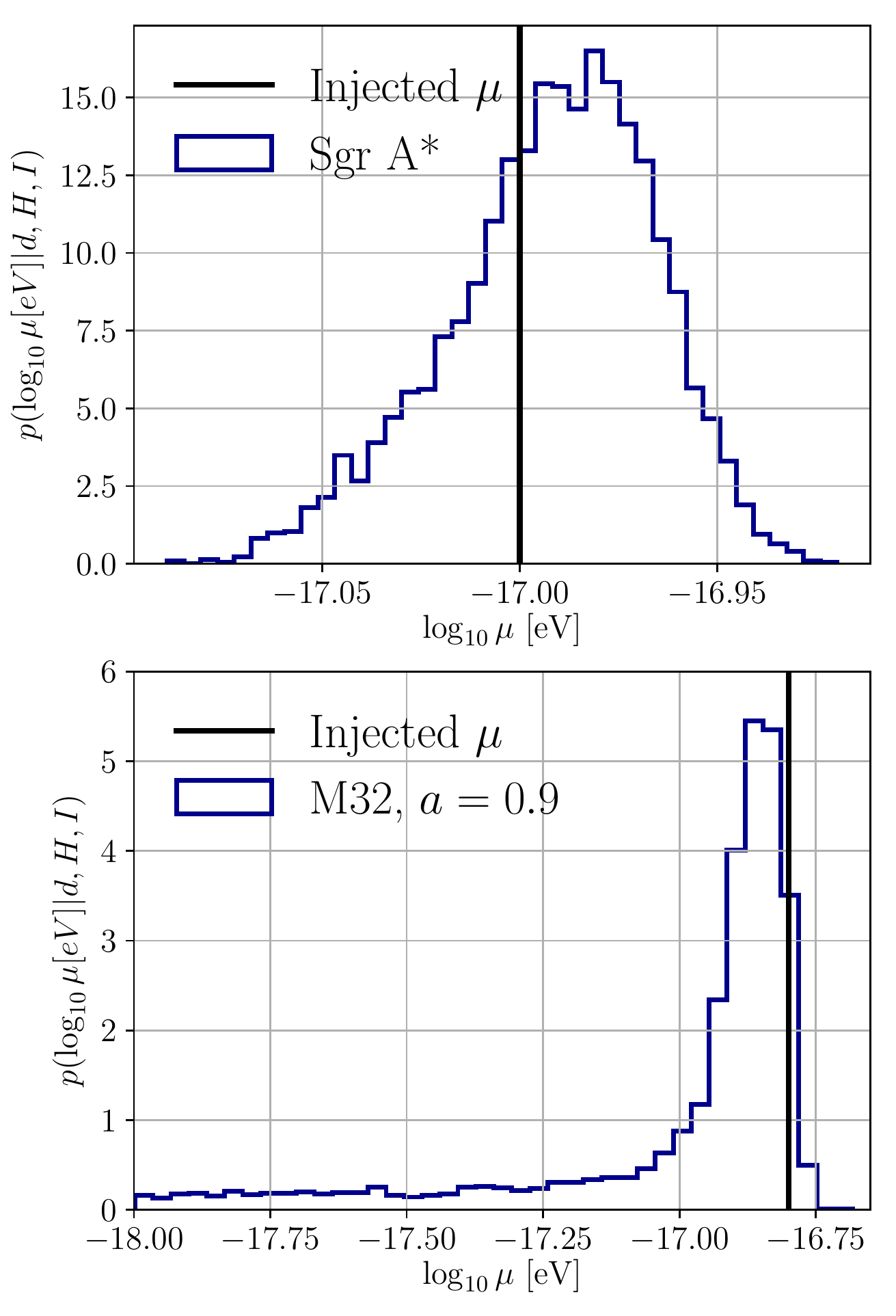}
 \caption{The marginalized posterior of $ \mu $ estimated from the same signal in \cref{fig:PDF_mu} except with an injected value of $\mu = 6.03 \times 10^{-18} \rm eV$ for Sagittarius A* (top panel) and $ \mu = 1.58 \times 10^{-17} \rm eV$ for M32 (bottom panel).
 For both injections, the injected $\mu$ corresponds to $M \mu \sim 0.3 $. 
 In both panels, the solid vertical line in black denotes the injected $\mu$.
 Both posteriors peak at a value of $\mu$ close to the injected value with significant support. 
 Moreover, the posterior of Sagittarius A* (solid blue line) peaks so much sharper than the posterior of M32 (solid red line) that it can only be seen clearly when zoomed in (embedded figure).
 The better measurement of $\mu$ for Sagittarius A* is expected because its relative proximity gives ringdown signals with higher SNR. 
 }
 \label{fig:PDF_mu_inj}
\end{figure}

Our constraint suggests that we might be able to probe the existence of ultralight boson of $\mu \sim 10^{-17} \rm eV $ by detecting just a single ringdown signal of these SMBHs. 
To test so, we inject the same ringdown signal of a black hole surrounded by a bosonic cloud. 
If $M \mu $ is too close to zero or unity, cloud formation is not favorable. 
Hence we assume $M \mu = 0.3 $ for M32 (corresponding to $\mu \sim 1.58 \times 10^{-17} \rm eV$) and Sgr A* (corresponding to $\mu \sim 6 \times 10^{-18} \rm eV$).
Based on \cite{Brito2017}, we assume the cloud-mass ratio to be $\alpha = 0.05 $, which is smaller than the maximum cloud mass that could develop around the assumed SMBHs. 
\cref{fig:PDF_mu_inj} shows the marginalized posteriors of $\log_{10} \mu $ obtained from the ringdown signal of Sgr A* (top panel) and M32 (bottom panel).
The posterior of both black holes peak at a value close to the injected $\mu$ (solid vertical line in black). 
The posterior for Sgr A* peaks more sharply at the injected $\mu$ because of its greater ringdown SNR.
These results indicate that our method can also recover an injected ultralight boson mass from solely detecting the ringdown waveform of a SMBH. 

\begin{figure*}[tp!]
 \includegraphics[width=\textwidth]{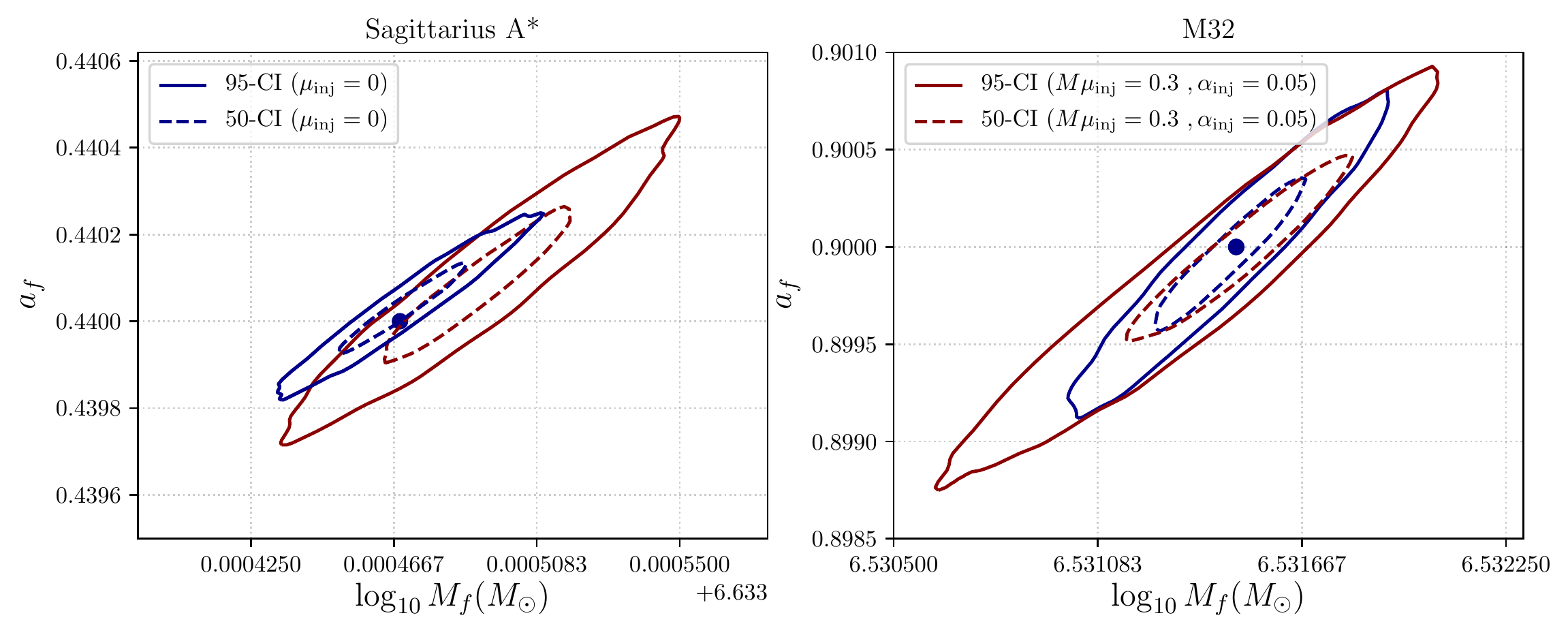}
 \caption{
 The 50\% (dashed) and 95\% (solid) CI of the two-dimensional posteriors of spin $a_f$ and the base-10 log of mass $\log_{10} M_f$ of Sgr A* (left panel) and M32 (right panel) with injected $\mu = 0 $ (blue) and $ M \mu = 0.3 $ (red). 
 The scatter points mark the mass and spin of the host black holes. 
 Contours on both panels enclose the respective injected values, 
meaning that while we measure $\mu$, we can also accurately measure the properties of the host black hole. 
 Thus, our analysis will not mistake a black hole with a bosonic cloud as a vacuum black hole of a different mass or spin. 
 }
 \label{fig:contour}
\end{figure*}

While we measure $\mu$, we can also accurately measure the mass and spin of the host black hole. 
\cref{fig:contour} shows the 50\% confidence interval (CI, solid lines) and 95\% CI (dashed lines) of the two dimensional posterior of host black-hole mass and spin recovered from our analysis of the mock signals from Sgr A* (left panel) and M32 (right panel). 
The scatter points on both panels mark the respective mass and spin of the black holes. 
As we can see, for the cases of $\mu = 0 $ (in blue) and $ M \mu = 0.3 $ (in red), all contours enclose the injected values, which are the assumed mass and spin of the host black holes. 
These show that we can accurately measure $M$, $a_f$ and $\mu$ simultaneously. 
These results also imply that our method will not mistake a black hole with a bosonic cloud as a vacuum black hole of different mass and spin. 

\section{Concluding Remarks}
\label{sec:conclusion}

In conclusion, we have shown a new method to calculate the shift of gravitational QNMFs due to a scalar field around a black hole. 
Focusing on scalar ultralight bosons, we calculated the QNMF shifts due to bosonic clouds surrounding the black hole. 
Using the computed frequency shift, we demonstrated that by a single detection of the ringdown phase EMRI signal of a nearby supermassive black hole, such as Sgr A* and M32, we can confirm or rule out the existence of ultralight bosons of mass between $ [10^{-17}, 10^{-16}] \rm eV $. 
Our method further extends the working scope of other proposed GW searches of ultralight bosons, which cover the range of $\mu \in [10^{-16}, 10^{-12}] \rm eV$ \cite{tsukada2019first, palomba2019direct, pani2012perturbations, ng2019searching, ng2020constraints, hannuksela2019probing}. 

To simplify our analysis, we have made use of several assumptions.
We ignored back reactions on the bosonic cloud during the inspiral and ringdown phase, which might cause depletion of the cloud, mass transfer between the cloud and the black holes, and the excitation of modes of the bosonic cloud as in  \cite{hannuksela2019probing, eda2013new}. 
This is an approximation which is similar to the Cowling approximation, which assumes a stationary background spacetime when studying the matter modes in asteroseismology. 
We note that the Cowling approximation should be valid in general \cite{Cowling_apprx_01, Cowling_apprx_02, Cowling_apprx_03, Cowling_apprx_04, Cowling_apprx_05, Cowling_apprx_06} but  \textbf{A2} might fail when studying some gravitational systems.
One example where neglecting effects of the perturbations on the background may break down is the calculation of quasinormal modes in massive Chern-Simon theory \cite{counter_example}.
However, in our case, the effect of cloud perturbations on the quasinormal modes should not be significant.
Firstly, in general, gravitational perturbations in the ringdown phase are weaker than those in the inspiral phase. 
For (extremely) small-mass-ratio inspirals, which our study focuses on, it has been verified that the excitation and decay of bosonic clouds during the inspiral phase are negligible \cite{baumann2019probing, berti2019ultralight}. 
Thus, the cloud excitation and decay during the ringdown phase should be even weaker compared to that of the inspiral phase. 
Secondly, in our study, the effective potential is proportional to $ (M \mu)^8 $. 
When $M \mu < 1$, the effective potential of the cloud is small relative to that of the Kerr background.
In contrast, in \cite{counter_example}, the effective potential due to scalar field is proportional to the coupling constant $\alpha$ and $\mu_{\rm CS}^2 $, where $\mu_{\rm CS}$ stands for the mass of the scalar field in massive Chern-Simons gravity. 
Thus, the gravitational feedback on bosonic clouds and the effects on the ringdown phase due to the feedback are more suppressed than massive Chern-Simons gravity. 
For these reasons, we expect our approximation is valid for our analysis.
Nonetheless, to prepare for future astrophysical detections, it will be beneficial to study the gravitational feedback of bosonic clouds more thoroughly with numerical efforts. 

Other than assuming the Cowling-like approximation (\textbf{A2}), we have also ignored any changes in the amplitudes of the quasinormal modes, focusing instead on the frequency shifts. 
Future in-depth studies to include the changes of amplitudes into our measurement could lead us towards probing lower boson masses at further distances, as effects of the bosonic cloud on quasinormal-mode excitation factors should lead to stronger ringdown waveform effects.
Moreover, when computing the shifts of QNMFs, we assume $ (M \mu)^2 \ll 1 $ (\textbf{A1}), which is a physically motivated regime for ultralight boson searches \cite{hannuksela2019probing, eda2013new, tsukada2020modeling}. 
We also note a few outstanding questions which may be of future interest: for one, whether the bosonic cloud around a black hole will break the isospectrality of gravitational quasinormal modes \cite{isospectralty} and lead to the emergence of new modes remain to be seen. 
We also hope to explore the effects of vector and tensor ultralight bosons on a black hole's gravitational quasinormal modes. 

Our work has several important implications. 
Firstly, our method is capable of probing the mass of ultralight bosons with ringdown analyses, both by constraining the mass in the absence of a detection, and via direct detection if the scalar ultralight boson actually exists.
Our ringdown-based analysis could bolster already existing inspiral-based analyses, and strengthen evidence for detection of ultralight bosons \cite{hannuksela2019probing}.
Secondly, our work illustrates a method for calculating shifts in the quasinormal modes due to surrounding environmental effects, which may be extended to other possible dark matter structures, like halos \cite{DM_halo_01, DM_spike_01, DM_spike_02, DM_spike_03} and spikes.
This sheds light on a whole new direction to detect signatures of dark matter. 
Thirdly, the potential derived in this paper may also affect the self-force calculations \cite{EMRI_Self_force_01, EMRI_Self_force_02, EMRI_Self_force_03, EMRI_Self_force_04, EMRI_Self_force_05, EMRI_Self_force_06} for estimating GWs generated by an EMRI, which is the skeleton of many proposed methods to search for the existence of ultralight bosons by dynamical friction. 
To fully account for the effects of a bosonic cloud on EMRI orbits, self-force calculations may need to consider the bosonic cloud effective potential we have derived. 

Other than searching for dark matter, our calculations may help to test scalar-field involved alternative theories, such as Gauss-Bonnet theories \cite{EdGB_01} and scalarized black holes \cite{BH_scalarization_01, BH_scalarization_02, BH_scalarization_03, BH_scalarization_04}. 
Existing studies of gravitational QNMFs of black holes in these theories are confined to nonrotating or slowly rotating black holes \cite{dCS_QNM_01, dCS_QNM_02, dCS_QNM_03, dCS_QNM_04, dCS_QNM_05, EdGB_QNM_01, EdGB_QNM_02}. 
This limitation hinders us from thoroughly testing these alternative theories from astrophysical black holes as they are usually spinning. 
We note that scalar fields described by energy-momentum tensors similar to \cref{eq:Stress-Tensor} appear in these theories. 
Calculations presented in this paper might be useful for extracting QNMFs of generically spinning black holes in these alternative theories.
Once these spectra are available, these theories can be subjected to more thorough GW tests, paving a new way to study environmental effects of black holes through GW detection. 

\section*{Acknowledgements}
The authors are indebted to Emanuele Berti and Otto A. Hannuksela for valuable discussion and their comments on the manuscript.
A.K.-W.C. was supported by the Hong Kong Scholarship for Excellence Scheme (HKSES). 
The work described in this paper was partially supported by grants from the Research Grants Council of the Hong Kong (Project No. CUHK 24304317), The Croucher Foundation of Hong Kong, and the Research Committee of the Chinese University of Hong Kong. 
This manuscript carries a report number of KCL-PH-TH/2021-47. 

\appendix

\begingroup
\allowdisplaybreaks
\begin{widetext}
\section{Explicit expressions for the effective potential and frequency shifts}
\label{sec:App_A}
The polynomials $C_n^{0 2 1}(x, y)$ and $C_n^{0 2 2}(x, y)$ in Eq.~\ref{eq:VULB} are 

\begin{align}
\begin{split}
    C_0^{0 2 1}(\tilde{r}, \tilde{k}) = & \, 4 \tilde{k}^2 \tilde{r}^5+\left(-16 \tilde{k}^2-8 \tilde{k}\right) \tilde{r}^4+\left(20 \tilde{k}^2+38 \tilde{k}+4\right) \tilde{r}^3+\left(-8 \tilde{k}^2-60 \tilde{k}-24\right) \tilde{r}^2\\
    &+\left(32 \tilde{k}+48\right) \tilde{r}-32\\
    C_2^{0 2 1}(\tilde{r}, \tilde{k}) = &-4 \tilde{k}^2 \tilde{r}^8+\left(14 \tilde{k}^2-4 \tilde{k}\right) \tilde{r}^7+\left(-24 \tilde{k}^2+22 \tilde{k}+2\right) \tilde{r}^6+\left(30 \tilde{k}^2-27 \tilde{k}-15\right) \tilde{r}^5\\
    &+\left(-24 \tilde{k}^2+12 \tilde{k}+30\right) \tilde{r}^4+\left(8 \tilde{k}^2-4 \tilde{k}-44\right) \tilde{r}^3+56 \tilde{r}^2-32 \tilde{r}\\
    C_1^{0 2 1}(\tilde{r}, \tilde{k}) =  & \, 0,
\end{split}
\end{align}

\begin{align}
\begin{split}
    C_0^{0 2 2}(\tilde{r}, \tilde{k}) = & -14 \tilde{k}^2 \tilde{r}^6+\left(68 \tilde{k}^2+14 \tilde{k}\right) \tilde{r}^5+\left(-118 \tilde{k}^2-87 \tilde{k}\right) \tilde{r}^4+\left(88 \tilde{k}^2+212 \tilde{k}+12\right) \tilde{r}^3\\
    &+\left(-24 \tilde{k}^2-236 \tilde{k}-72\right) \tilde{r}^2+\left(96 \tilde{k}+144\right) \tilde{r}-96\\
    C_2^{0 2 2}(\tilde{r}, \tilde{k}) = & \, 18 \tilde{k}^2 \tilde{r}^7+\left(14 \tilde{k}-60 \tilde{k}^2\right) \tilde{r}^6+\left(90 \tilde{k}^2-103 \tilde{k}-16\right) \tilde{r}^5+\left(-72 \tilde{k}^2+148 \tilde{k}+108\right) \tilde{r}^4\\
    &+\left(24 \tilde{k}^2-68 \tilde{k}-216\right) \tilde{r}^3+224 \tilde{r}^2-96 \tilde{r}\\
    C_1^{0 2 2}(\tilde{r}, \tilde{k}) =  & \, 0.
\end{split}
\end{align}

The polynomials $D_n^{0 2 1}(x)$ and $D_n^{0 2 2}(x)$ in Eq.~\ref{eq:Delta} are
\begin{align}
\begin{split}
   D_0^{0 2 1} (\tilde{k}) = & -256 \tilde{k}^8+704 \tilde{k}^7-160 \tilde{k}^6+288 \tilde{k}^5\\
   D_2^{0 2 1} (\tilde{k}) = & \, 2048 \tilde{k}^8-7680 \tilde{k}^7+9184 \tilde{k}^6-11168 \tilde{k}^5+7432 \tilde{k}^4-3672 \tilde{k}^3+864 \tilde{k}^2\\
   D_4^{0 2 1} (\tilde{k}) = & -6144 \tilde{k}^7+22720 \tilde{k}^6-22464 \tilde{k}^5+18584 \tilde{k}^4-7484 \tilde{k}^3+1092 \tilde{k}^2+336 \tilde{k}\\
   D_6^{0 2 1} (\tilde{k}) = & \, 7680 \tilde{k}^6-27136 \tilde{k}^5+20464 \tilde{k}^4-11164 \tilde{k}^3+1830 \tilde{k}^2+402 \tilde{k}-24\\
   D_8^{0 2 1} (\tilde{k}) = & -5120 \tilde{k}^5+17360 \tilde{k}^4-9268 \tilde{k}^3+2980 \tilde{k}^2-23 \tilde{k}-15\\
   D_{10}^{0 2 1} (\tilde{k}) =& \, 1920 \tilde{k}^4-6432 \tilde{k}^3+2214 \tilde{k}^2-379 \tilde{k}+16\\
   D_{12}^{0 2 1} (\tilde{k}) = & -384 \tilde{k}^3+1380 \tilde{k}^2-274 \tilde{k}+30\\
   D_{14}^{0 2 1} (\tilde{k}) = & \, 32 \tilde{k}^2-160 \tilde{k}+16\\
   D_{16}^{0 2 1} (\tilde{k}) = & \, 8\\
   D_j^{0 2 1} (\tilde{k}) =&\, 0 \quad \quad \text{for odd} \quad j,
\end{split}\\[3ex]
\begin{split}
   D_0^{0 2 2} (\tilde{k}) = & \, 1768 \tilde{k}^8+576 \tilde{k}^7-1536 \tilde{k}^6+1936 \tilde{k}^5-1008 \tilde{k}^4\\
   D_2^{0 2 2} (\tilde{k}) = & \, 6144 \tilde{k}^8-18432 \tilde{k}^7+19872 \tilde{k}^6-12896 \tilde{k}^5+3368 \tilde{k}^4+72 \tilde{k}^3\\
   D_4^{0 2 2} (\tilde{k}) = & -18432 \tilde{k}^7+44736 \tilde{k}^6-46144 \tilde{k}^5+24360 \tilde{k}^4-4100 \tilde{k}^3-588 \tilde{k}^2\\
   D_6^{0 2 2} (\tilde{k}) = & \, 23040 \tilde{k}^6-45312 \tilde{k}^5+41488 \tilde{k}^4-15628 \tilde{k}^3+550 \tilde{k}^2+498 \tilde{k}\\
   D_8^{0 2 2} (\tilde{k}) = & -15360 \tilde{k}^5+22800 \tilde{k}^4-18556 \tilde{k}^3+4462 \tilde{k}^2+236 \tilde{k}-48\\
   D_{10}^{0 2 2} (\tilde{k}) =& \, 5760 \tilde{k}^4-5184 \tilde{k}^3+4274 \tilde{k}^2-603 \tilde{k}-14\\
   D_{12}^{0 2 2} (\tilde{k}) = & -1152 \tilde{k}^3+36 \tilde{k}^2-470 \tilde{k}+44\\
   D_{14}^{0 2 2} (\tilde{k}) = & \, 96 \tilde{k}^2+192 \tilde{k}+20\\
    D_{16}^{0 2 2} (\tilde{k}) = & \, -24\\
   D_j^{0 2 2} (\tilde{k}) =&\, 0 \quad \quad \text{for odd} \quad j.
\end{split}
\end{align}

The functions $E^{0 2 1}(x)$ and $E^{0 2 2}(x)$ in Eq.~\ref{eq:Delta} are
\begin{align}
    E^{0 2 1}(\tilde{k},\tilde{\mu},\tilde{b}) = &\left(24 \tilde{k} \tilde{\mu }^2-14 \tilde{k}^2-5 \tilde{k}-8 \tilde{\mu }^4-2 \tilde{\mu }^2\right) \text{Ei}(2 \tilde{b})+4 i \pi  \tilde{b}^2,\\
     E^{0 2 1}(\tilde{k},\tilde{\mu},\tilde{b}) = &\left(-24 \tilde{k} \tilde{\mu }^2-6 \tilde{k}^2+\tilde{k}+12 \tilde{\mu }^4+20 \tilde{\mu }^2\right)\left(\text{Ei}(2 \tilde{b})-i \pi\right),
\end{align}
where $\text{Ei}(x)$ is the exponential integral special function
\begin{equation}
    \text{Ei}(x) = \int^{x}_{- \infty} d z \frac{e^{z}}{z}.\\
\end{equation}
\end{widetext}
\endgroup
When deriving the potentials, we have made use of the following fact which we found numerically, 
\begin{equation}
\begin{split}
& \oint d \Omega S_{022} Y_{22}(\theta, \phi) \sim \oint d \Omega S_{021} Y_{21}(\theta, \phi)  \sim 1. 
\end{split}
\end{equation}

\section{Calculations of logarithmic perturbation}
\label{sec:App_B}

To calculate the QNMFs due to an additional, perturbative potential, we apply the method of logarithmic perturbations. 
To start, it will be more convenient for us to transform \cref{eq:radial_TE_w_source} into a Klein-Gordon-like equation, by letting $ u = \Delta^{-1} \sqrt{r^2 + a^2} R(r) $  \cite{Nakamura_01}. 
Then $ u $ satisfies, 
\begin{equation}\label{eq:TE_tarsnformed}
\frac{\partial^2 u}{\partial x^2} + \left( \omega^2 - V^{(0)} (x) - V^{\rm KG}_{\rm ULB}(r) \right) u = 0, 
\end{equation}
where $ x $ is the tortoise coordinate defined by $ \frac{d}{d x}=\frac{\Delta}{r^{2}+a^{2}} \frac{d}{d r} $, $ V^{(0)}(x) $ is the intrinsic potential, and $ V^{\rm KG}_{\rm ULB}(r) $ is the transformed potential due to bosonic cloud, which is related to $ V^{\rm (ULB)} $ by 
\begin{equation}
V^{\rm KG}_{\rm ULB} = - \frac{\Delta}{(r^2 + a^2)^2} V_{\rm ULB}(r). 
\end{equation}
As $ (M \mu)^8 \ll 1 $, $ |V^{\rm KG}_{\rm ULB}| \ll | V^{(0)}|$, and by logarithmic perturbation theory, we can compute the QNMFs up to the leading order of $ \mathcal{O}(V^{\rm KG}_{\rm ULB}) $. 
Formally, the shifted QNMFs are given by \cite{LPT}
\begin{equation}
\tilde{\omega}_{n \ell m} \approx \tilde{\omega}^{(0)}_{n \ell m} + \frac{\braket{u|V^{\rm KG}_{\rm ULB}|u}}{2 \tilde{\omega}^{(0)}_{n \ell m} \braket{u|u}}, 
\end{equation}
where $ \tilde{\omega}^{(0)} $ is the QNMFs for $ \mu = 0 $, which we take to be the values in \cite{Berti_QNM}, $ u $ is the solution to Eq.~\ref{eq:TE_tarsnformed} when $ V^{\rm KG} = 0 $,  and
\begin{equation}\label{eq:LPT_integral}
\begin{split}
\braket{u|u} &=\int_{-\infty}^{+\infty} d x u^{2}(x), \\
\braket{u|V^{\rm KG}_{\rm ULB}(x)|u}& = \int_{-\infty}^{\infty} d x u(x) V^{\rm KG}_{\rm ULB}(x) u(x). 
\end{split}
\end{equation}
Since $u(x) \sim e^{ikx}$, for a negative imaginary component of the QNMF $\bra{u}V_{\rm ULB}\ket{u}$ is dominated by contributions from the far field.
In this limit, 
\begin{equation}
R(r) \propto r^3 e^{+i \omega r}. 
\end{equation}
Thus, in the far-field limit, we can make the following approximations to simplify our calculations of $ \tilde{T} $, 
\begin{equation}
\begin{split}
\partial_r R(r) &= +i \omega r^3 e^{+i \omega r} + 3 r^2 e^{+i \omega r} \approx i \omega R(r), \\
\partial_2^2 R(r) &\approx - \omega^2 r^3 e^{+i \omega r} = - \omega^2 R(r). 
\end{split}
\end{equation}
These approximations are valid because $ r/M \gg 1 $ and the terms of the highest power of $ r $ dominate. 
Therefore, we take
\begin{equation}
u(r) \propto r^2 e^{+i \omega r}. 
\end{equation}
Evaluating Eq.~\ref{eq:LPT_integral}, we have 
\begin{equation}\label{eq:QNM_LPT_02}
\tilde{\omega}_{n \ell m} \approx \tilde{\omega}_{n \ell m}^{(0)} + \frac{\Delta_{n \ell m}(\mu, M_s, M)}{2 \tilde{\omega}_{n \ell m}^{(0)}}. 
\end{equation}

\bibliographystyle{apsrev4-1}
\bibliography{sample}

\end{document}